\documentclass{pic2012}
\usepackage{epstopdf}
\def\runauthor{A.A.Savin}
\def\shorttitle{Production of the W and Z bosons at hadron colliders}
\begin{document}

\title{
Production of W and Z bosons at hadron colliders
}

\author{Alexander A. Savin}

\address{University of Wisconsin\\
1150 University Ave., Madison, USA\\
on leave from \\ Skobeltsyn Institute of Nuclear Physics, Moscow State University, Moscow, Russia\\
E-mail: asavin@mail.cern.ch }

\maketitle

\abstracts{
The article summarizes the main recent measurements related to
production of the W and Z bosons at the Tevatron and the LHC experiments.
The results of the measurements are compared to the standard model
predictions.
}

\section{Introduction} 

The measurements of the W and Z bosons production in hadron-hadron collisions 
provide an important test of the standard model (SM) of particle physics.
Study of the W and Z production is also important, since this is a major
source of background for 
searches for new physics beyond SM.
The theoretical predictions of the inclusive W and Z cross 
sections are available at next-to-next-to-leading order (NNLO) in perturbative QCD. 
The calculations are limited by uncertainties on parton distribution functions (PDFs), 
higher-order QCD, and electroweak radiative corrections, which are available at 
next-to-leading order (NLO).

This
article gives a short overview of the most recent measurements from 
the Tevatron experiments
(CDF and D0), with the total integrated luminosity of $\rm{p\bar{p}}$ collisions
at $\sqrt{s}=1.96$ TeV of up to 10.5 fb$^{-1}$, and from the LHC experiments 
(ATLAS, CMS, LHCb), with
approximately 5 fb$^{-1}$ of pp collisions collected at each 7 and 8 TeV 
center-of-mass energies.

\section{Inclusive W and Z production}
\subsection{Measurements of
inclusive W and Z production cross sections}

The most recent measurement of the
inclusive W and Z production
cross sections was performed at the LHC in pp collisions collected in 
2012 at $\sqrt{s} = 8$ TeV and corresponding to an integrated luminosity of 
$18.7 \pm 0.9$ pb$^{−1}$~\cite{CMS-PAS-SMP-12-011}. 
The LHC instantaneous luminosity increased dramatically compared to the 
dataset used for the first measurements at 
7 TeV~\cite{JHEP-10-2011-132,PR-D85-2012-072004}, from 
$2 \times 10^{31}$ to $7 \times 10^{33} \rm{cm}^{-2}\rm{s}^{-1}$, and 
the average number of inelastic proton-proton interactions (pileup) 
increased
 from 
two up to twenty.
Since the precision measurement requires
a low pileup and triggers with low transverse momentum thresholds,
the instantaneous luminosity was decreased by approximately a factor of ten
 in a dedicated LHC configuration.
The LHC beams were separated in
the transverse plane to diminish the effective overlap. The working point was 
chosen to allow manageable rates with an unprescaled single 
lepton trigger with sufficiently loose requirements, and at the same 
time yielding a low average pileup. 
\begin{figure}[htbp]
\begin{center}
\includegraphics[width=0.45\textwidth]{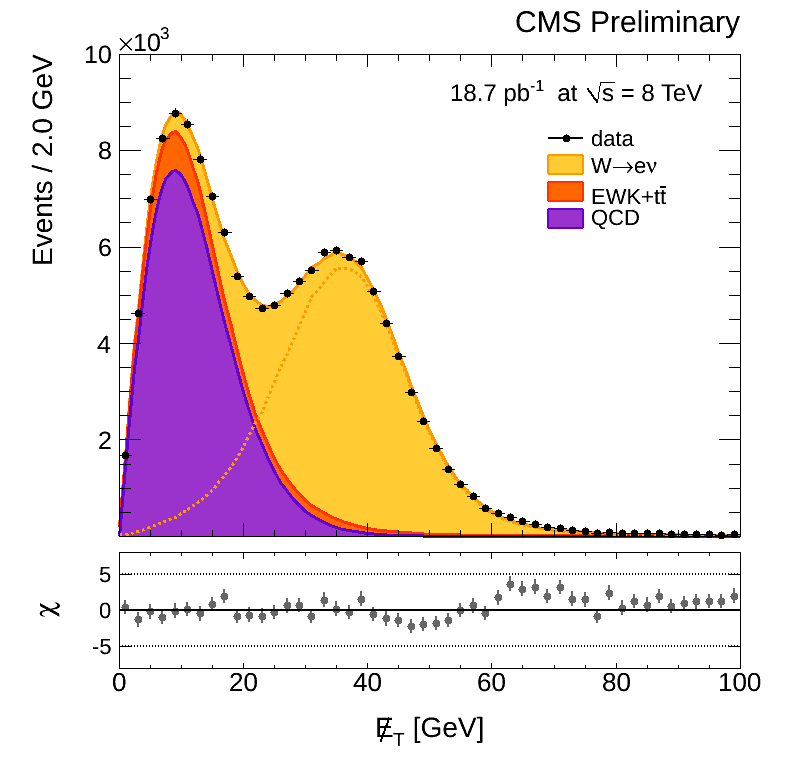}
\includegraphics[width=0.45\textwidth]{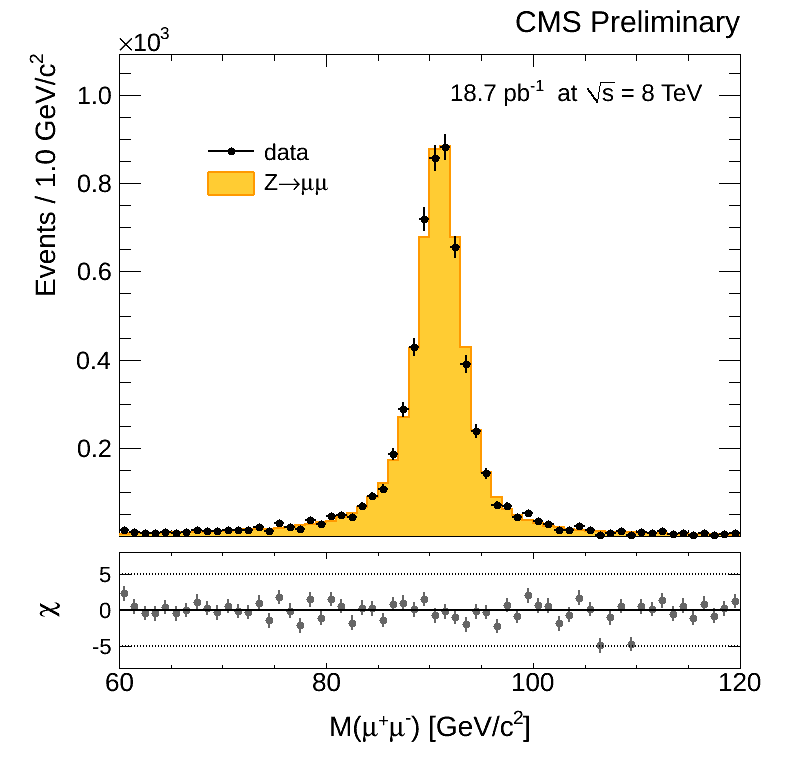}
\caption{ (left) The
$E_{\rm{T}}^{\rm{miss}}$ distribution for the selected $\rm{W} \rightarrow e\nu$
candidates. (right) The dilepton mass distributions for the selected
$\rm{Z} \rightarrow \mu\mu$ candidates.
\label{fig:WZincl}}
\end{center}
\end{figure}

Leptonic W boson decays are characterized by a prompt, energetic, and 
isolated lepton and a
neutrino giving rise to significant missing transverse energy, 
$E_{\rm{T}}^{\rm{miss}}$. The $E_{\rm{T}}^{\rm{miss}}$ distribution for the 
selected $\rm{W} \rightarrow e\nu$ events is presented in Fig.~\ref{fig:WZincl}(left).
The Z boson decays to leptons are selected based on two energetic and isolated leptons. 
The reconstructed dilepton mass is used to estimate efficiencies and 
measure event yields,
and is presented in Fig.~\ref{fig:WZincl}(right) for the $\rm{Z} \rightarrow \mu\mu$ decay
channel.

\begin{figure}[htbp]
\begin{center}
\includegraphics[width=0.45\textwidth]{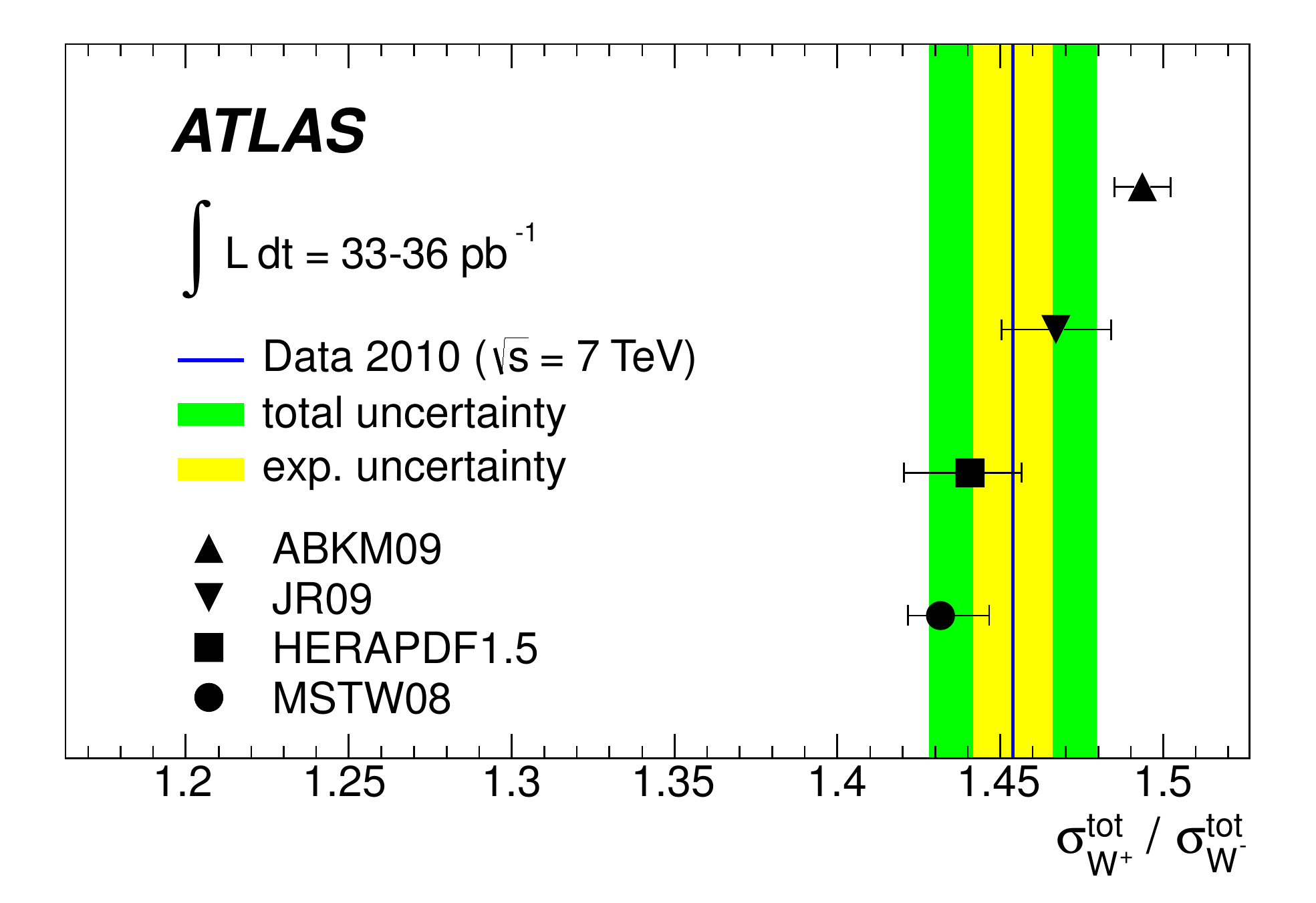}
\includegraphics[width=0.45\textwidth]{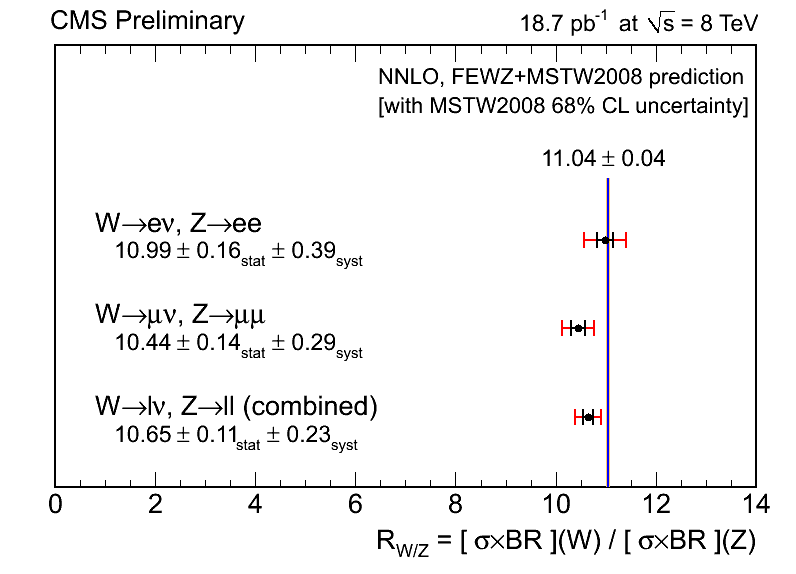}
\caption{ 
Summary of the measurements of the ratios of 
$\rm{W}^+$ 
to $\rm{W}^-$ and 
W to Z  
production cross sections. 
Measurements in the electron and muon channels are combined and compared to the 
theoretical predictions computed at the NNLO with different PDF sets. 
Left plot presents the ATLAS measurements at 7 TeV~\protect\cite{PR-D85-2012-072004}
 and the right plot the CMS
measurements at 8 TeV~\protect\cite{CMS-PAS-SMP-12-011}. 
\label{fig:WZratios}}
\end{center}
\end{figure}

\begin{figure}[htbp]
\begin{center}
\includegraphics[width=0.7\textwidth]{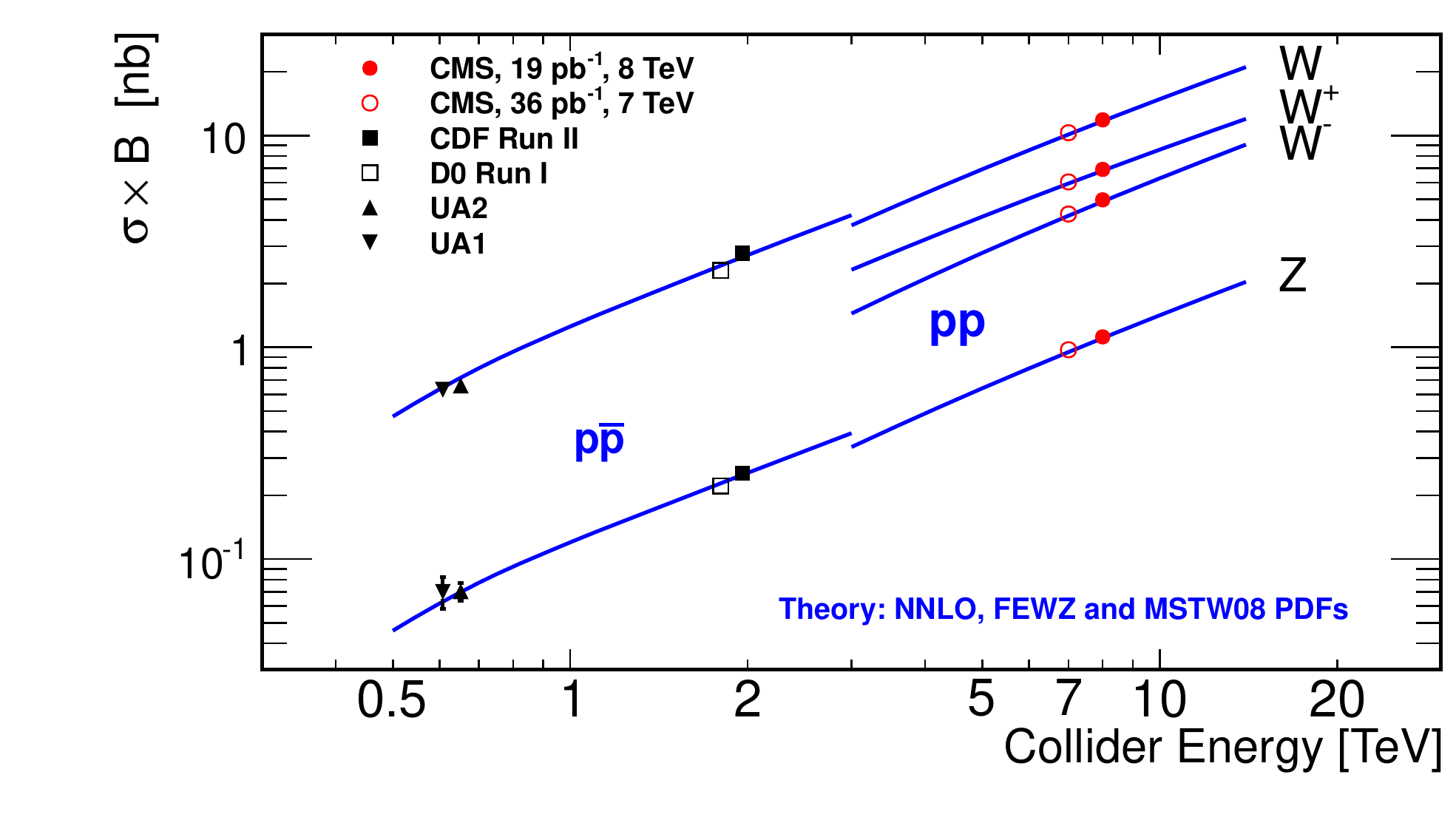}
\caption{ 
Products of inclusive W and Z production cross sections and branching 
ratios as a function of center-of-mass energy~\protect\cite{CMS-PAS-SMP-12-011}.
The lines are the theoretical predictions at NNLO.
\label{fig:XSasS}}
\end{center}
\end{figure}

Assuming lepton universality, the measurements in the different lepton decay modes are combined
by calculating an average value weighted by the combined statistical and systematic 
uncertainties, taking into account the correlated uncertainties. 
The most powerful way to compare the measured values and theoretical predictions are
the ratios of cross sections, where some of the systematic uncertainties cancel. 
Figure~\ref{fig:WZratios} presents the 
summary of the CMS measurements of the ratios of the W to Z production  cross sections 
 and ATLAS measurements at 7 TeV
of the $\sigma_{\rm{W}^+}/\sigma_{\rm{W}^-}$. 
The measurements at both 7 and 8 TeV agree well with the standard model predictions.
The 7 and 8 TeV measurements are compared to the results of the low-energy experiments
in Fig.~\ref{fig:XSasS}. The ATLAS measurements at 7 TeV are
not presented but they agree well with the corresponding CMS points. The NNLO
predictions describe the data in the whole center-of-mass range.

\subsection{Study of differential distributions}

\begin{figure}[htbp]
\begin{center}
\includegraphics[width=0.37\textwidth]{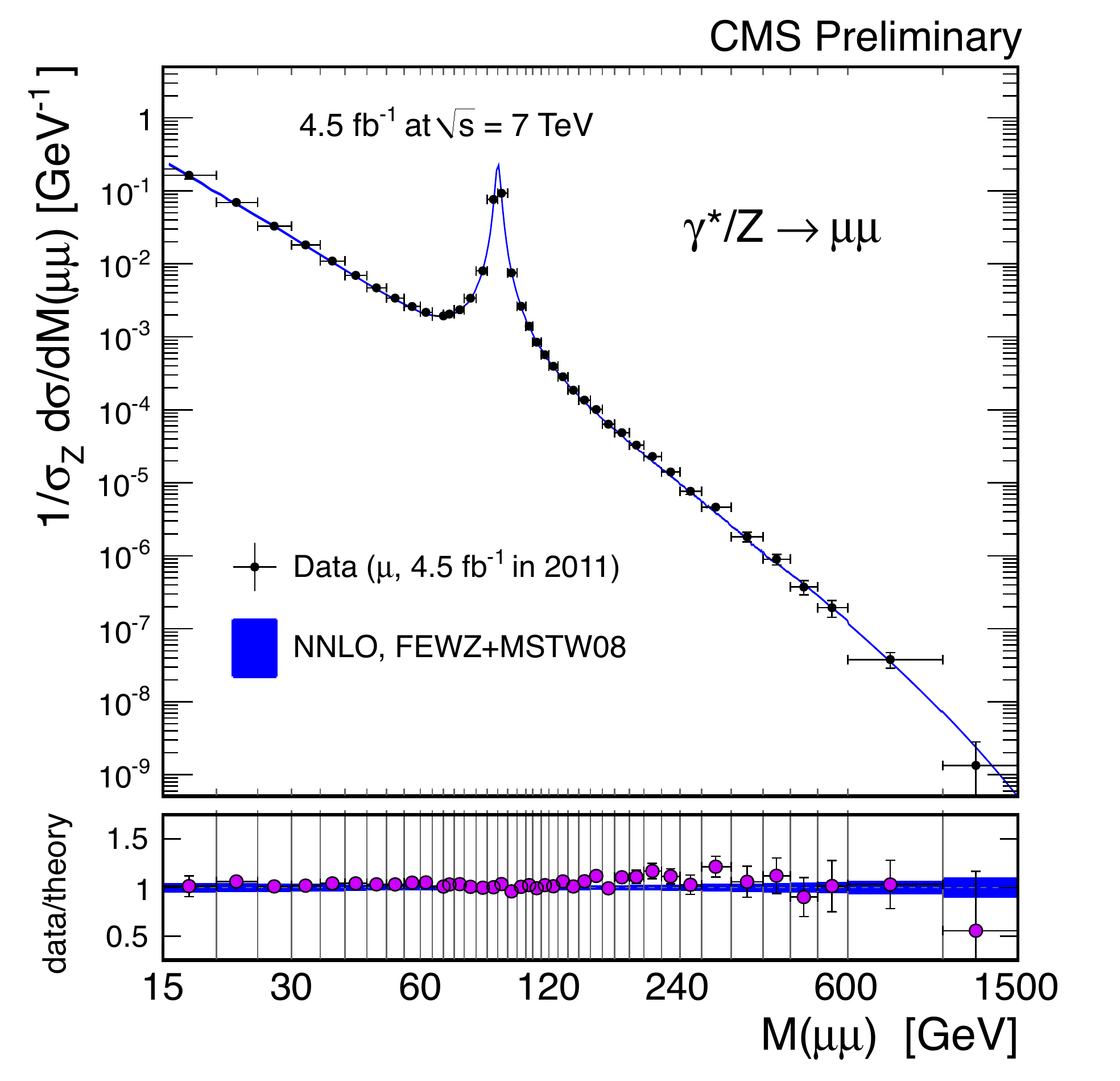}
\includegraphics[width=0.54\textwidth]{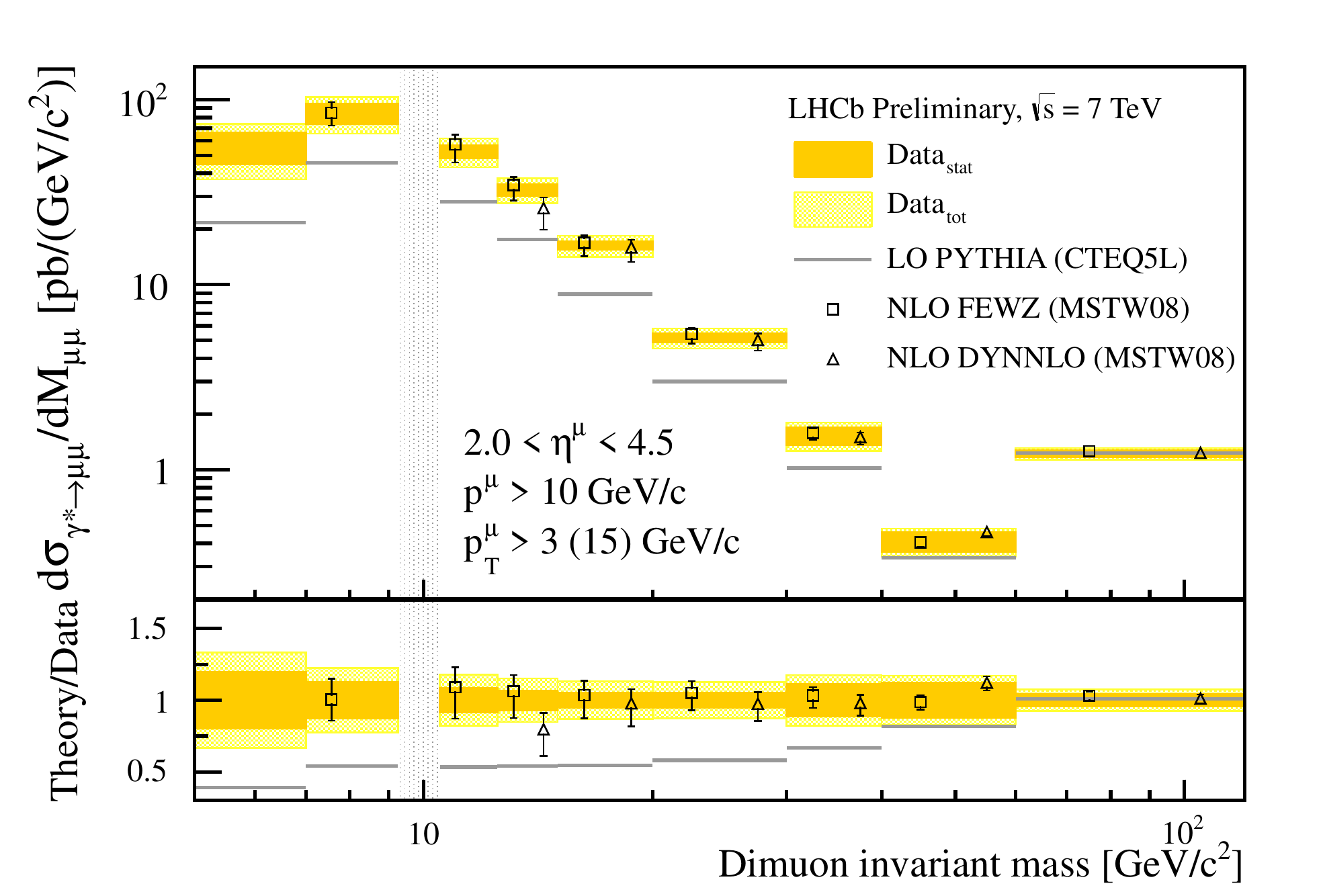}
\caption{ (left) The
 Drell-Yan invariant mass spectrum, normalized to the Z resonance region,
as measured by CMS and as predicted by NNLO calculations, (right)
as measured by LHCb and compared to LO, NLO  and NNLO predictions.
\label{fig:dymm}}
\end{center}
\end{figure}

\begin{figure}[htbp]
\begin{center}
\includegraphics[width=0.45\textwidth]{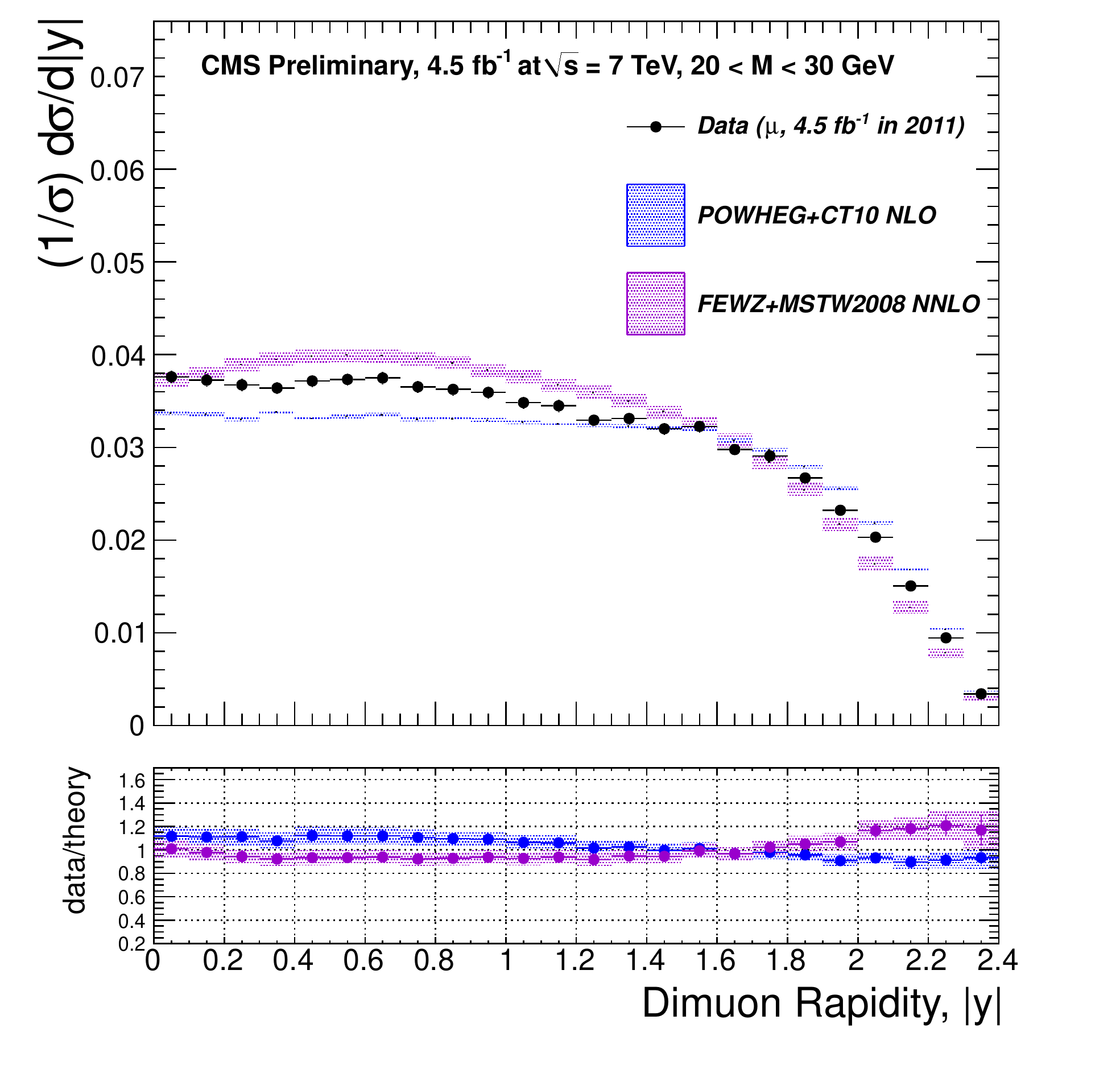}
\includegraphics[width=0.45\textwidth]{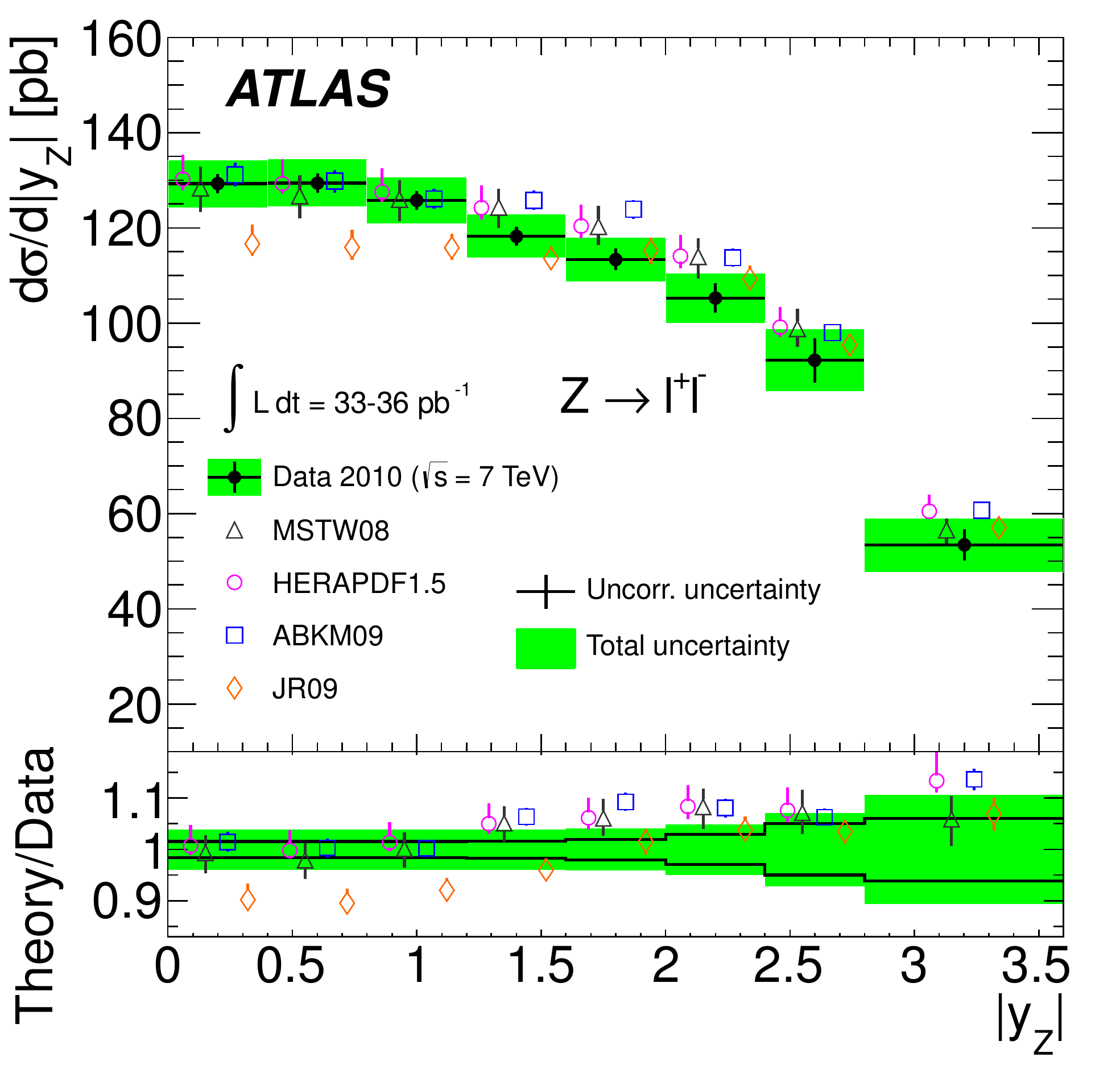}
\caption{ (left) The
Drell-Yan cross section
$(1/\sigma)d\sigma/d|\rm{Y}|$ as a function of dimuon rapidity, measured in 
the dilepton mass range 20--30 GeV at 7 TeV by CMS and
(right) as measured by the ATLAS experiment in the region 66--116 GeV.
\label{fig:dymmdy}}
\end{center}
\end{figure}

Theoretical calculations of the differential cross section 
of the Drell-Yan (DY) production, $d\sigma/d{\rm{M}}$, 
and the double-differential cross section, 
$d\sigma/d{\rm{M}}d{\rm{Y}}$,
 where M is the dilepton 
invariant mass, and Y is the absolute value of the dilepton rapidity
 are well established up to the NNLO. 
Comparisons between theoretical calculations and experimental measurements 
provide important 
constrains on the PDFs. 
The measurements presented in Fig.~\ref{fig:dymm} are
performed with CMS dataset consisting of 4.5 fb$^{-1}$ of 
proton-proton
collision data in the dimuon channel collected 
at the LHC at a centre-of-mass energy of
$\sqrt{s} = 7$ TeV~\cite{CMS-PAS-EWK-11-007}.
The measurements are normalized to the Z peak region (60 - 120 GeV), 
which cancels out the uncertainty on the integrated luminosity and 
reduces the PDF uncertainty on acceptance, the pileup effect in the 
reconstruction efficiency and the uncertainty on the efficiency estimation. 
The NNLO predictions describe the mass spectrum generally well. The 
right plot presents LHCb measurements~\cite{LHCb-CONF-2012-013} and effect of using different orders of
theoretical calculations on the final cross section.
In the double
differential cross section presented in Fig.~\ref{fig:dymmdy}(left)
 at low dilepton masses 
the agreement between data and theory becomes pure. 
The difference in predictions
due to different PDFs, as shown in Fig.~\ref{fig:dymmdy}(right), 
makes the measurement
important for the future PDF constrains.

\subsection{Measurements of the W charge asymmetry and polarization}

\begin{figure}[htbp]
\begin{center}
\includegraphics[width=0.5\textwidth]{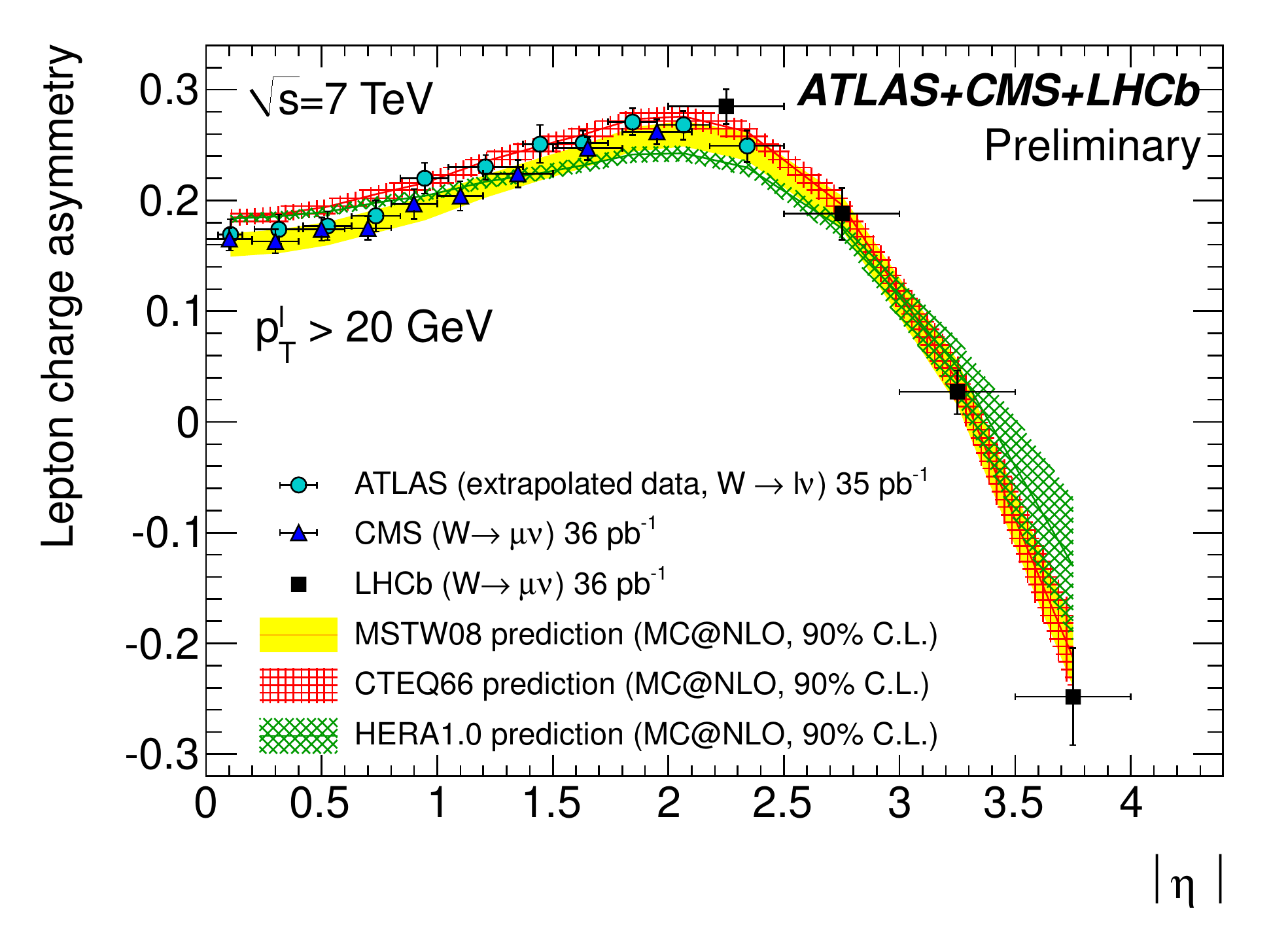}
\includegraphics[width=0.4\textwidth]{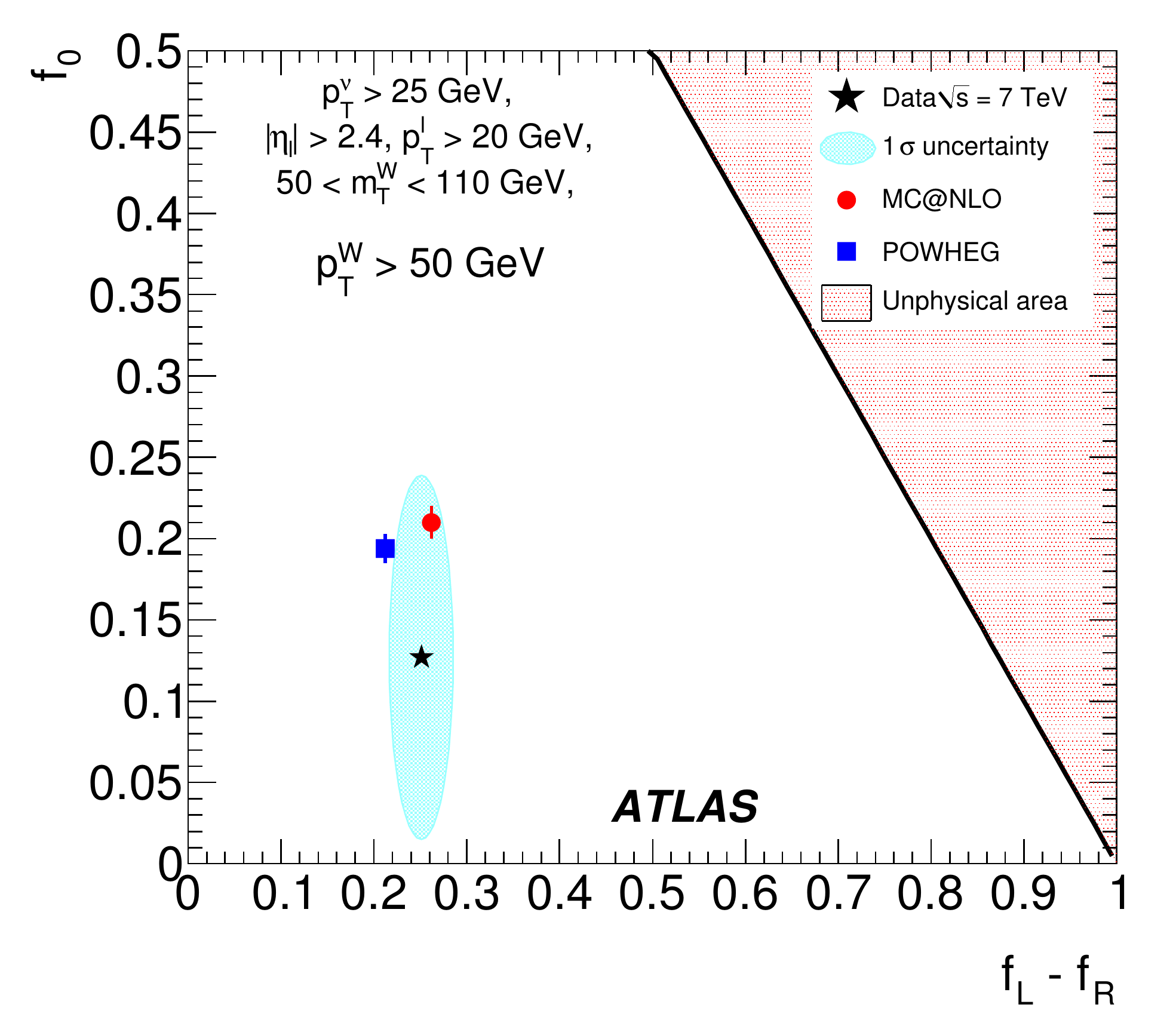}
\caption{ (left) The
lepton charge asymmetry from W boson decays as a function of absolute 
pseudorapidity measured at the ATLAS, CMS, and LHCb 
experiments~\protect\cite{ATLAS-CONF-2011-129}.
(right) Measured values of $f_0$ and $(f_L-f_R)$ for $P_T^{\rm{W}} > 50$ GeV, 
compared with the theoretical predictions~\protect\cite{EPJ-C72-2012-2001}.
\label{fig:WChAss}}
\end{center}
\end{figure}

The W boson charge asymmetry is mainly sensitive to valence quark distributions via the
dominant production process $u\bar{d}(\bar{u}d) \rightarrow \rm{W}^{+(-)}$. In pp collisions the
production rate of W$^+$ bosons is significantly larger than the corresponding W$^-$ rate.
The asymmetry is defined as
\begin{equation}
A_\ell = \frac {d\sigma_{\rm{W}_{\ell^+}}/d\eta_\ell - d\sigma_{\rm{W}_{\ell^-}}/d\eta_\ell}
 {d\sigma_{\rm{W}_{\ell^+}}/d\eta_\ell + d\sigma_{\rm{W}_{\ell^-}}/d\eta_\ell} 
\nonumber
\end{equation}
Figure~\ref{fig:WChAss} shows the lepton charge asymmetry from W-boson decays 
in bins of absolute pseudorapidity for the three LHC experiments. 
While the ATLAS data show the extrapolated combined asymmetry for muon 
and electron measurements, the LHCb and CMS measurements have been performed 
using only the $\rm{W} \rightarrow \mu\nu$ decay channel. In this plot the 
maximum of the asymmetry around $|{\eta_\ell}| \approx 2.0$ can be seen very
 well as the turn-over in asymmetry, which takes negative values 
above $|{\eta_\ell}| \approx 3.0$. This behaviour is related to the kinematics 
of the V-A decay of the W boson, its polarisation and the different 
fractions of the W$^+$ and W$^-$ bosons produced via high-$x$ sea quarks.
The measured asymmetry is well described by the predictions with different PDFs.
At the LHC the quarks generally carry a larger fraction of the momentum of the 
initial-state protons than the antiquarks. This causes the W bosons to be boosted 
in the direction of the initial quark. In the massless quark approximation, the quark 
must be left-handed and the antiquark right-handed. 
For more centrally produced W bosons, there is an increasing probability 
that the antiquark carries a larger momentum fraction than the quark, so the 
helicity state of the W bosons becomes a mixture of left- and right-handed 
states whose proportions are respectively described with fractions $f_L$ and $f_R$.
At high transverse momenta more complex production mechanisms contribute, and 
polarisation in longitudinal states, $f_0$, is also possible. 
This state 
is particularly interesting as it is directly connected to the massive character 
of the gauge bosons.
The measured values of $f_0$ and $(f_L-f_R)$ from~\cite{EPJ-C72-2012-2001}
 are presented in Fig.~\ref{fig:WChAss}(right) 
together with the NLO theoretical predictions.
The results show that predictions reproduce well the relative fractions of the 
left- and right-handed
states, but the data 
favour lower values of $f_0$ than predicted. Similar CMS measurements can 
be found in~\cite{PRL-107-2011-021802}. 

\subsection{Measurements of the DY forward-backward asymmetry and effective weak mixing angle}

\begin{figure}[htbp]
\begin{center}
\includegraphics[width=0.48\textwidth]{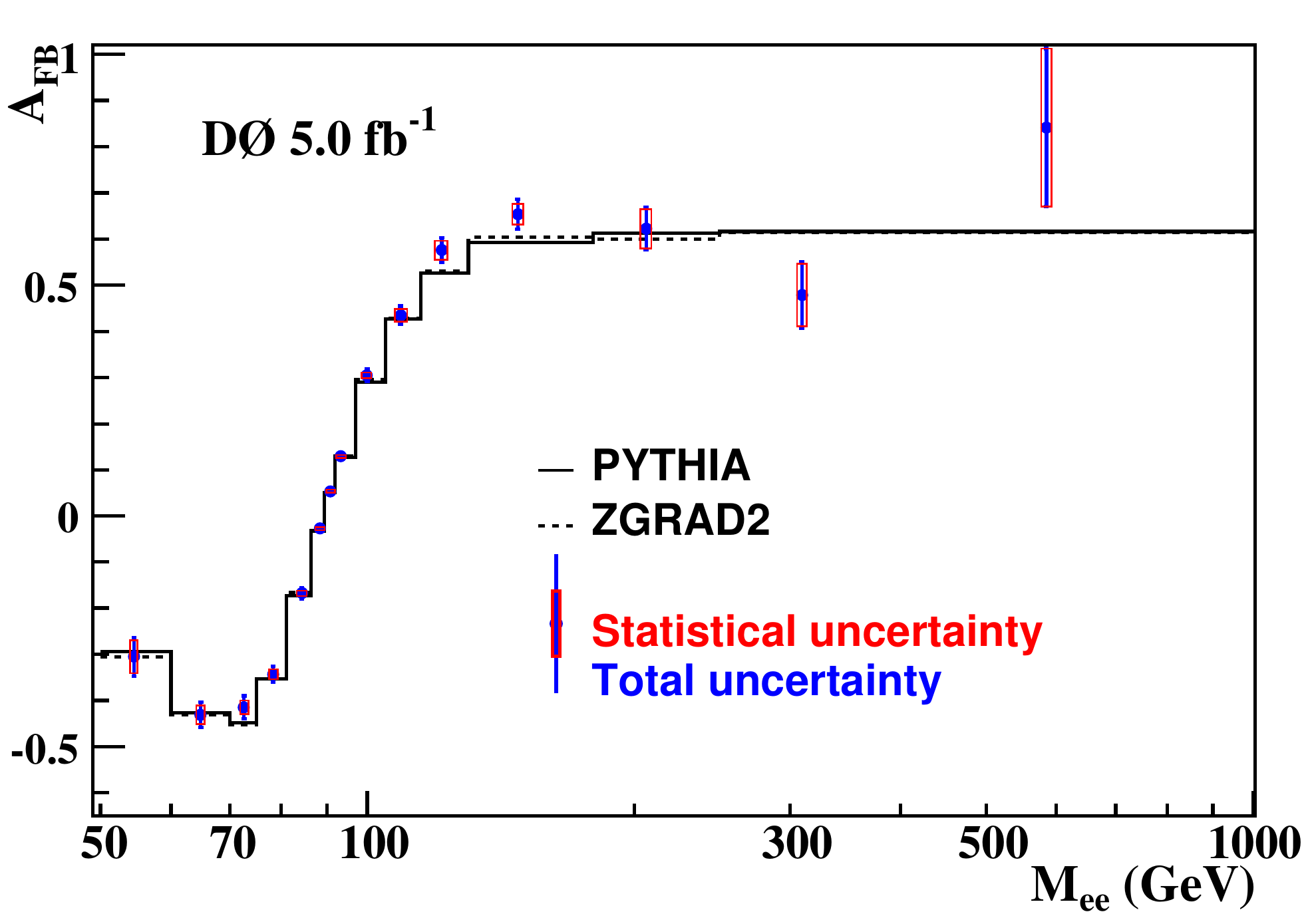}
\includegraphics[width=0.42\textwidth]{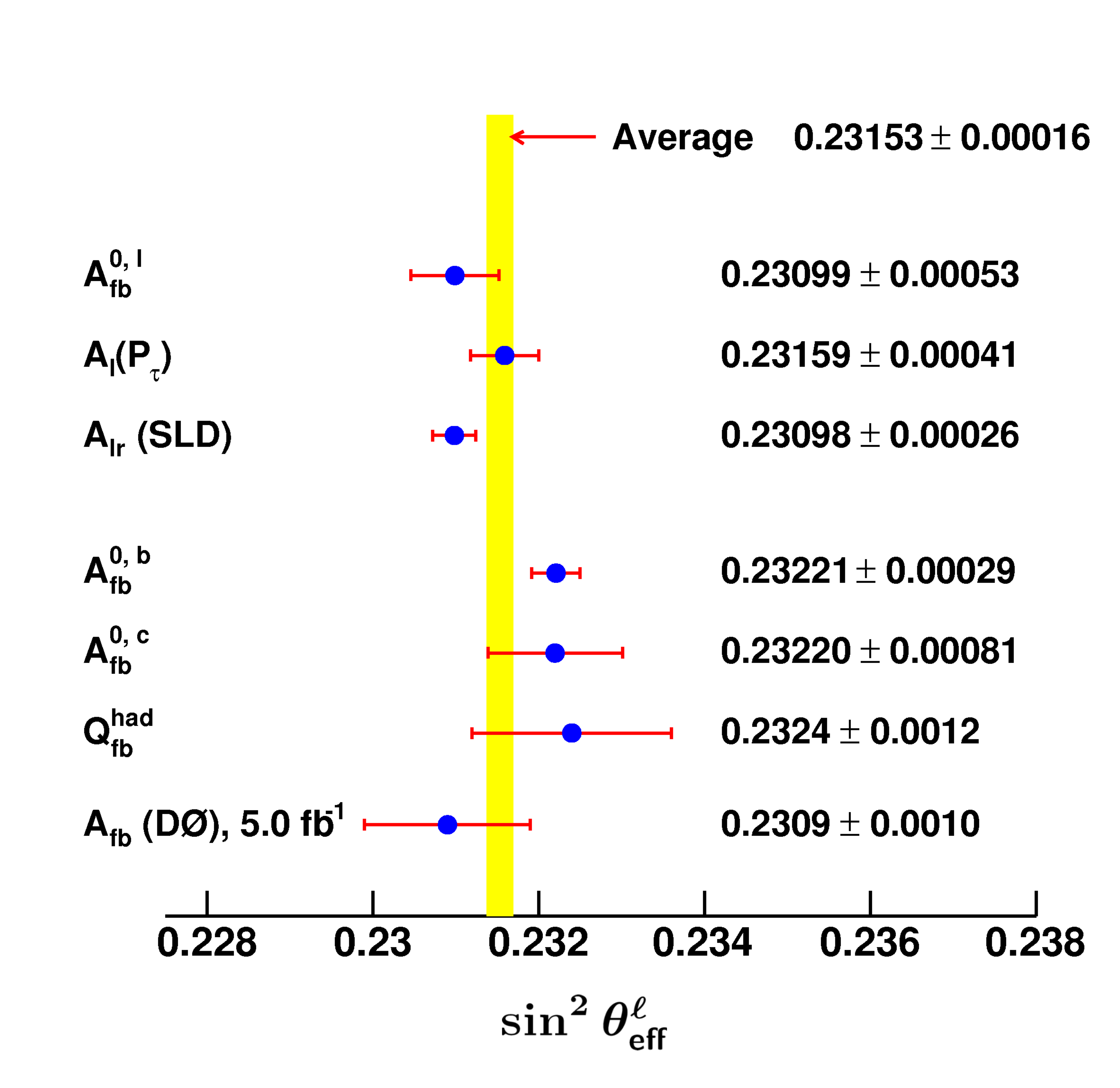}
\caption{ (left) 
Comparison between the measured $A_{FB}$ (points) and 
theoretical predictions (solid and dashed curves).
The boxes and vertical lines show 
the statistical and total uncertainties, respectively.
(right)
Comparison of measured $sin^2\theta_{eff}^f$ from
different experiments. 
\label{fig:ForwBackw}}
\end{center}
\end{figure}

In the SM, the DY production occurs to first order via 
$q\bar{q}$ annihilation into a real (or virtual) Z boson
or a virtual photon ($\gamma^*$). 
While the coupling of a fermion to the photon is purely a vector coupling, 
the coupling of the same fermion to the Z boson has both vector 
and axial-vector components.
The presence of both vector and axial-vector couplings gives 
rise to an asymmetry in the distribution of the polar angle $\theta^*$ of 
the negatively charged lepton relative to the incoming quark 
direction in the rest frame of the lepton pair. 
Events with $cos(\theta^*) > 0$ are classified as forward (F), 
and those with $cos(\theta^*) < 0$ are classified as backward (B). 
The forward-backward charge asymmetry, $A_{FB}$, is defined by
\begin{equation}
A_{FB}=\frac {\sigma_F - \sigma_B} {\sigma_F + \sigma_B} 
\nonumber
\end{equation}
where $\sigma_F$ and $\sigma_B$ are the cross sections for 
forward and backward processes, respectively.
The measured by D0~\cite{PRD-84-2011-012007} and predicted values of $A_{FB}$ are compared in Fig.~\ref{fig:ForwBackw}(left)
as a function of the dilepton mass.
Around the Z pole, the asymmetry is proportional to both the vector 
and axial-vector couplings of the Z boson to the fermions and 
is numerically close to 0. At large invariant mass, the asymmetry is dominated 
by Z/$\gamma^*$ interference and is almost constant ($\approx 0.6$). 
In the high mass region, the
measurement can be used to investigate possible new phenomena that may 
alter $A_{FB}$, such as new neutral gauge bosons or large extra dimensions, 
which are not yet observed in data.
In the vicinity of the Z pole, $A_{FB}$ is sensitive to the effective weak mixing 
angle $sin^2\theta_{eff}^f$ for each fermion
species ($f$) involved in a particular measurement. The comparison on the measured
in~\cite{PRD-84-2011-012007} $sin^2\theta_{eff}^f$ value to those measured in other experiments is
shown in Fig.~\ref{fig:ForwBackw}(right). The recent DY forward-backward asymmetry
measurement at CMS~\cite{PRD-84-2011-112002}
also showed no evidence for new physics at high masses. 
The week mixing angle was 
measured to be $0.229 \pm 0.020 (stat.) \pm 0.025 (sys.)$ in a good agreement
with the previous measurements. 

\section{Diboson production}
  
Many extensions of the SM predict new scalar,
vector, or spin-2 particles that decay into a pair of W or Z bosons. 
In addition, these
final states are sensitive to the self-interactions among the gauge bosons via trilinear
gauge couplings (TGCs). 
The values of these couplings are fully determined in the SM by the gauge structure
of the Lagrangian. 
The presence of anomalous neutral trilinear couplings (ATGCs) 
would lead to a sizable signal enhancement in
the cross sections via $s$-channel $q\bar{q}$ scattering.

\subsection{W$\gamma$ and Z$\gamma$ production}

The cross sections of the W$\gamma$ and Z$\gamma$ processes are often  measured 
as a function of the photon $E_T^{\gamma}$ threshold. 
Assuming lepton universality for the W and Z boson decays, the measured 
cross sections in the electron and muons channels are combined to reduce the 
statistical uncertainty. 
The MCFM~\cite{MCFM} program is commonly used to predict the NLO cross section. 
It includes photons from direct 
diboson production, from final state radiation off the leptons in 
the W/Z decays and from quark/gluon fragmentation into an 
isolated photon. Possible effects of composite W and Z boson 
structure can be simulated through the introduction of ATGCs.
\begin{figure}[htbp]
\begin{center}
\includegraphics[width=0.4\textwidth]{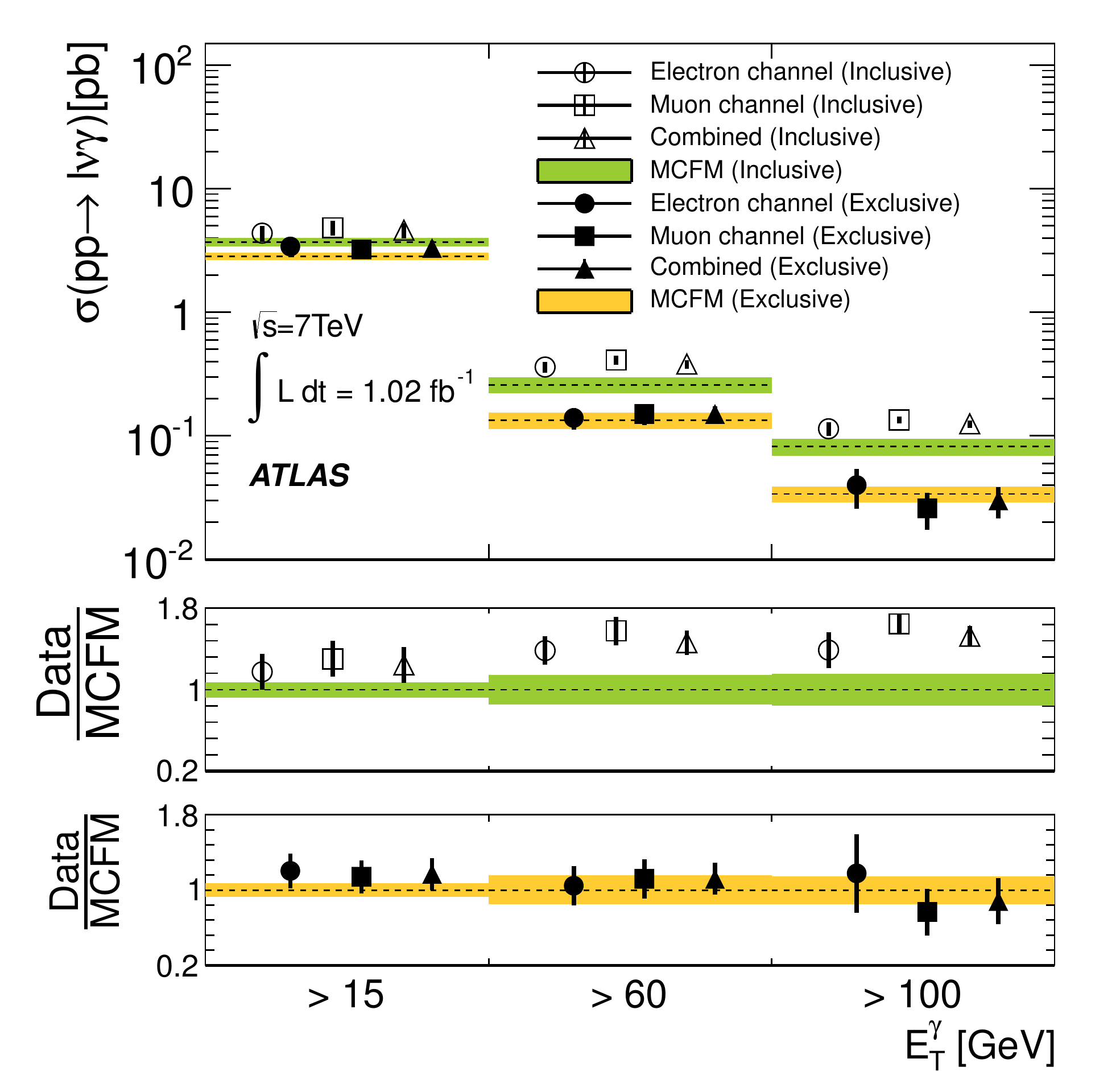}
\includegraphics[width=0.5\textwidth]{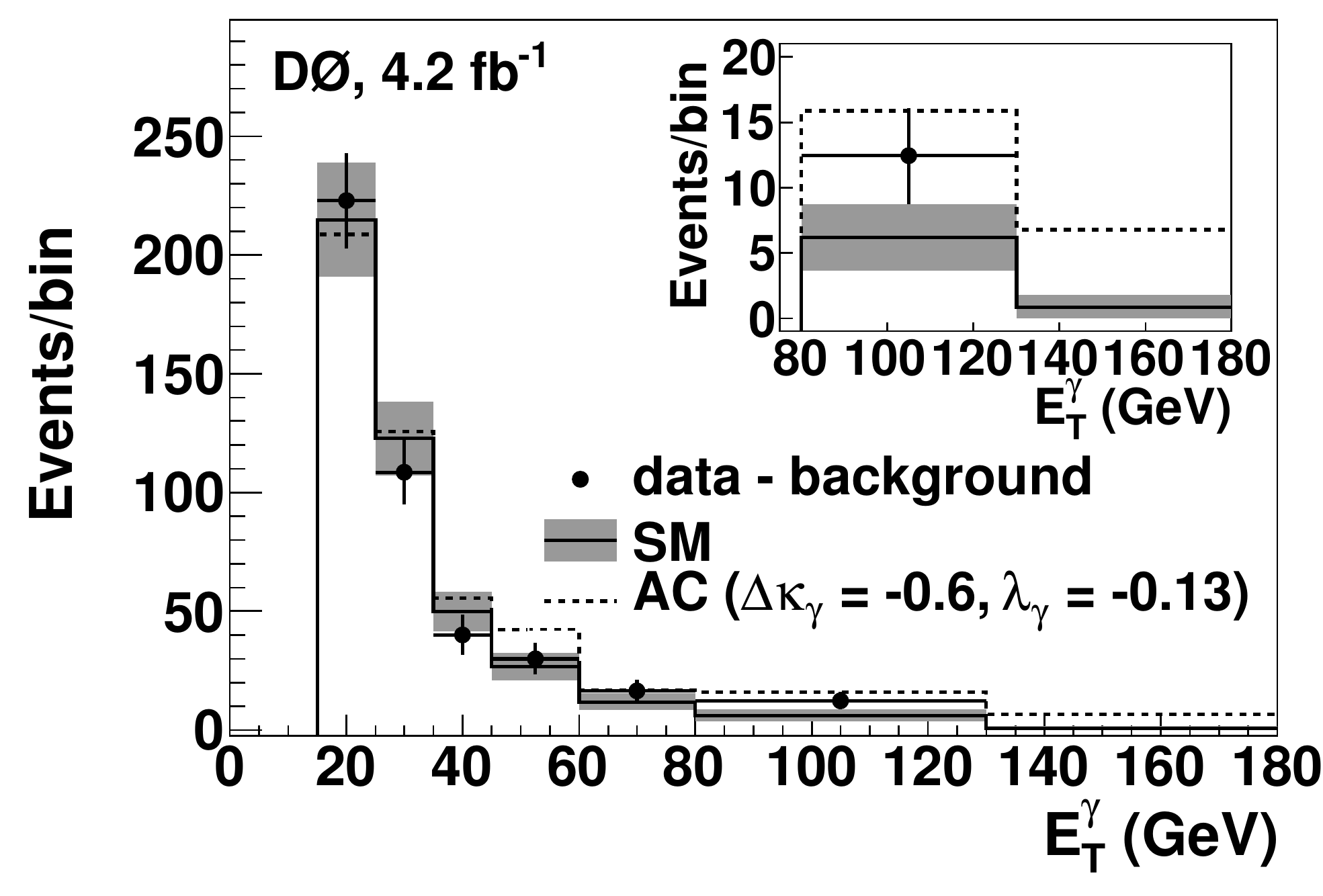}
\caption{ (left) 
The measured cross section for W$\gamma$ production compared to 
 the SM predictions. The measurements are performed 
in different $E_T^\gamma$ and jet multiplicity regions. 
(right)
Photon transverse energy distributions measured in data compared to 
the expectation for the SM and for one choice of ATGCs. The data are shown as 
black points.
The predictions are given by the solid (SM) 
and dashed (including ATGC) lines. The last $E_T^{\gamma}$ bin shows 
the sum of all contributions above 130 GeV. 
\label{fig:Wgamma}}
\end{center}
\end{figure}
The recent measurements from ATLAS~\cite{arXiv:1205.2531} quoted inclusive 
cross section, 
using only the lepton and photon selection cuts, and exclusive, requiring 
no jet with $E_T > 30$ GeV in the final state.
Figure~\ref{fig:Wgamma}  presents a summary of W$\gamma$ production measurements
 made in this study and the corresponding SM expectations. There is a good 
agreement between the measured cross sections for the exclusive events and the MCFM prediction.
For inclusive production, the MCFM NLO cross section prediction includes 
real parton emission processes only up to one radiated quark or gluon. 
The lack of higher-order QCD contributions results in an 
underestimation of the predicted cross sections, especially for 
events with high-$E_T$ photons, which have significant contributions from multijet final states.

The spectra of high energy photons in W$\gamma$ and Z$\gamma$
events are sensitive to new phenomena that alter the
couplings among the gauge bosons. These effects can be
described by modifying the WW$\gamma$ coupling $\kappa_\gamma$ 
from its
SM value of one and adding terms with new couplings.
The photon $E_T^\gamma$ distribution in Fig.~\ref{fig:Wgamma}(right)
shows a good agreement between data~\cite{PRL-107-2011-241803} and the SM predictions and used to derive limits 
on WW$\gamma$ ATGC using a binned likelihood fit to data. 
The 95\% CL limits on the WW$\gamma$ coupling parameters  
$−0.4 < \Delta\kappa_\gamma < 0.4$ and $−0.08 < \lambda_\gamma < 0.07$ were 
obtained by setting 
one coupling parameter to the SM value and allowing the other to vary.
More information can also be found 
in~\cite{PRD-85-2012-052001,PRL-107-2011-051802}.

\subsection{WW production}

\begin{figure}[htbp]
\begin{center}
\includegraphics[width=0.45\textwidth]{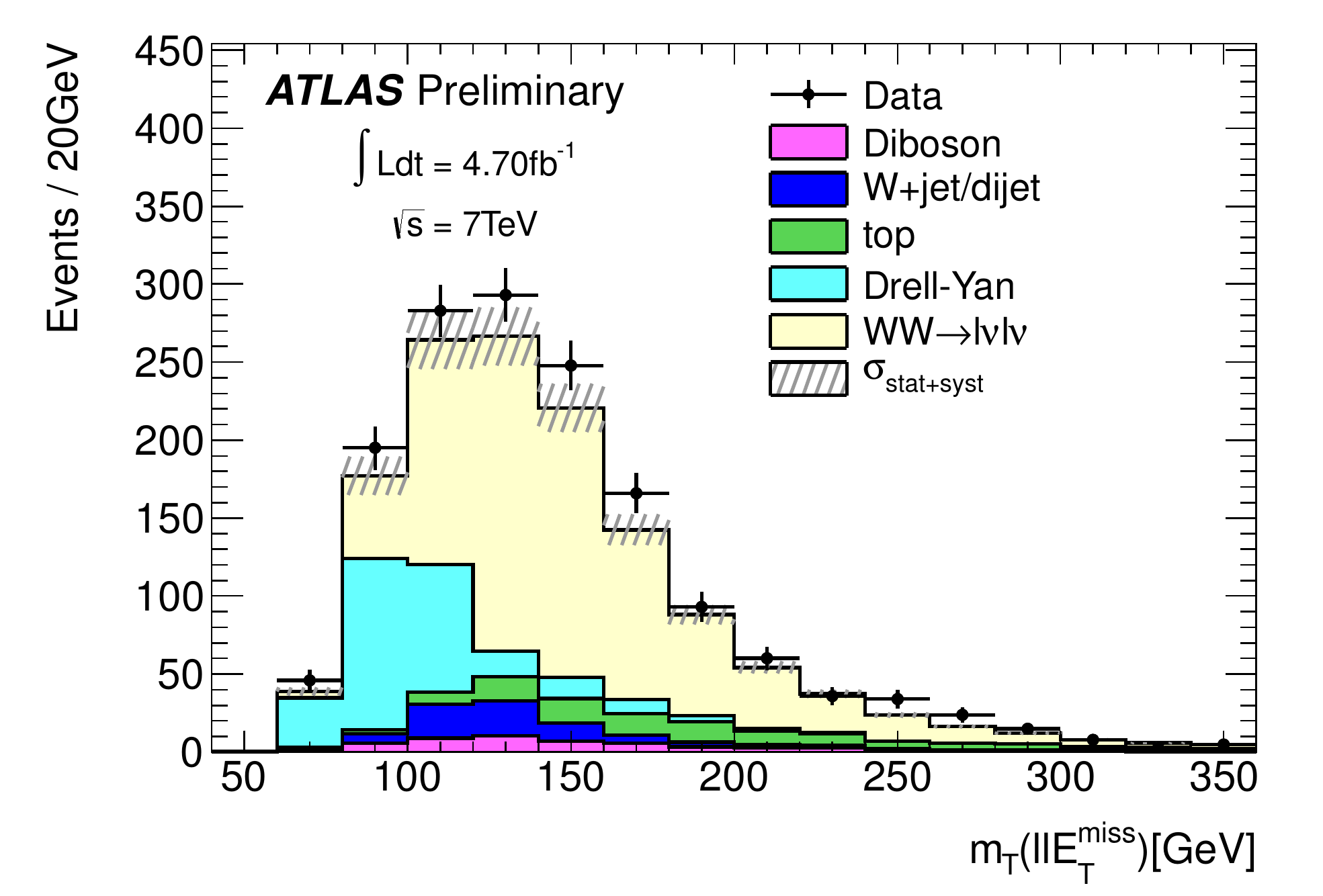}
\includegraphics[width=0.45\textwidth]{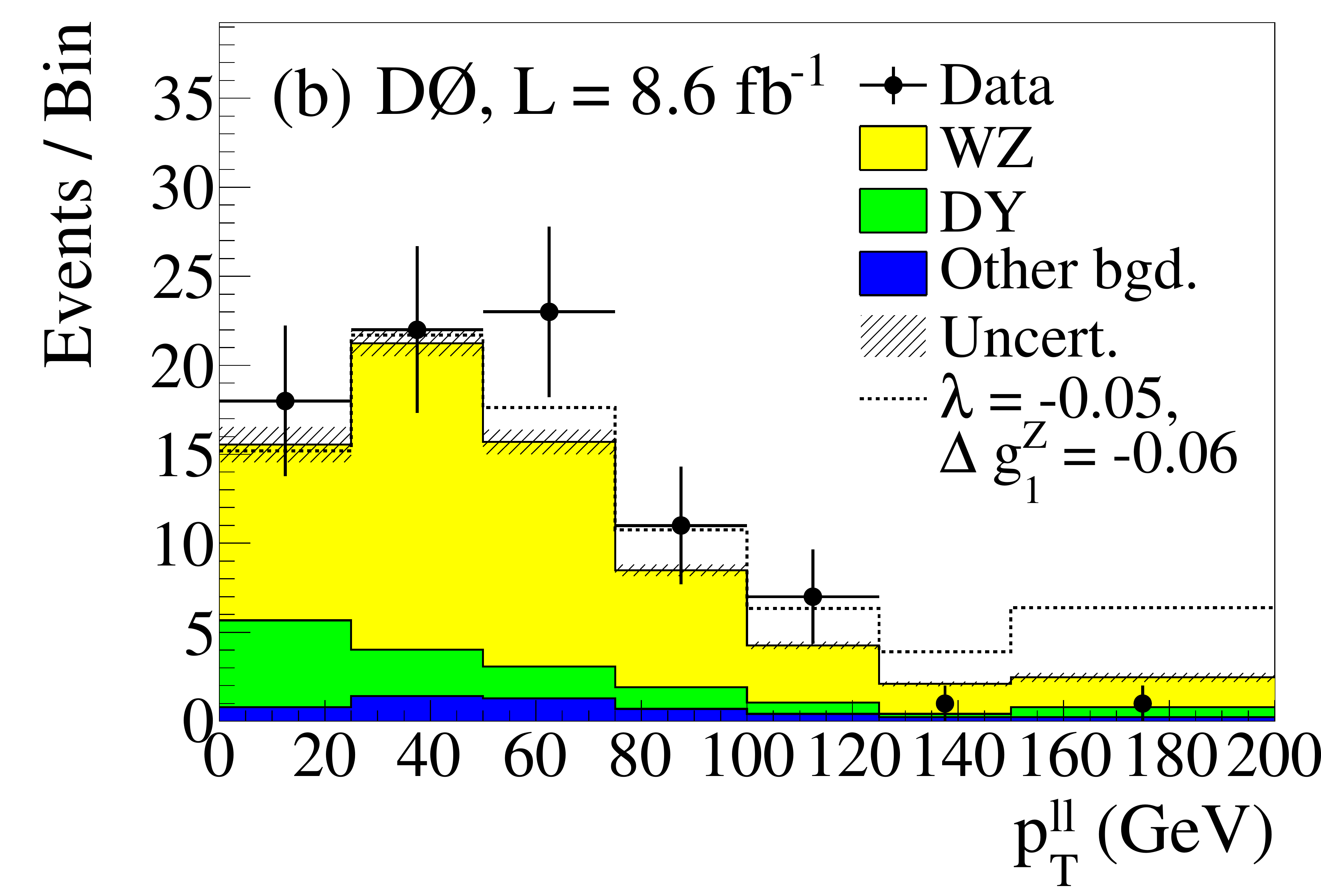}
\caption{ (left) 
Distributions of $m_T^{\ell\ell,{E_T^{miss}}}$ for the 
WW candidates after the combination of all channels. The points represent 
data and the stacked histograms the signal expectation and background 
estimates. 
(right)
The $p_T^{\ell\ell}$
distribution summed over all channels from $\rm{WZ} \rightarrow \ell\nu\ell\ell$
production for data, SM MC predictions and for ATGC 
model with $\lambda = −0.05$ and $\Delta g^1_Z = −0.06$.
\label{fig:WWandWZ}}
\end{center}
\end{figure}

A measurement of the WW production, which has a higher cross section than
other diboson channels, suffers also from high background contributions.
In 1524 observed $\rm{WW} \rightarrow \ell\nu\ell\nu$ candidate events in 
4.7 fb$^{-1}$ of
pp collisions at LHC,  ATLAS estimated 531 background 
events~\cite{ATLAS-CONF-2012-025}.
After applying selections on leptons, the dominant contribution ($>99\%$) to 
$ee$ and $\mu\mu$ events comes from the inclusive Z production. 
To reduce this background and the background contribution from 
hadronic multijets, 
the invariant mass of any dilepton pair is required to exceed 15 GeV and is 
not allowed to be within $\pm$15 GeV of the Z mass. 
Further suppression of background events is achieved by  applying a cut on 
the minima $E_T^{miss}$.
Suppression of the top contribution 
is achieved  by rejecting events containing more than 1 jet with $P_T > 25$ GeV and $|\eta| < 4.5$, or containing b-tagged jets.
The distributions of transverse mass 
of the dilepton and $E_T^{miss}$ system after all the selections 
is shown in Fig.~\ref{fig:WWandWZ}(left), where
\begin{equation} 
m_T^{\ell\ell,{E_T^{miss}}} = 
\sqrt{2 P_T^{\ell\ell} E_T^{miss} (1-\cos\Delta\phi)}\ ,
\nonumber
\end{equation}
where
$\Delta\phi$
is the difference in azimuth between $E_T^{miss}$ and $P_T^{\ell\ell}$.
The data and the predicted distributions are in good agreement.

Both ATLAS and CMS~\cite{PAS-CMS-SMP-12-005,PAS-CMS-SMP-12-013} measurements
of the WW cross section agree well with the SM predictions. Most recent
CMS measurement at 8 TeV~\cite{PAS-CMS-SMP-12-013}, 
$69.9 \pm 2.8(stat.)\pm 5.6(syst.) \pm 3.1(lumi.)$ pb, can be compared with
$57.3^{+2.4}_{-1.6}$ pb predicted by MCFM.

\subsection{WZ production}

The WZ cross section measurements performed by CDF and D0  showed a good agreement 
with the SM expectations. The CDF measurement~\cite{arXiv:1202.6629} of
$3.93 ^{+0.60}_{-0.53} (stat.) ^{+0.59}_{-0.46} (syst.)$ pb agrees well with   
$3.21 \pm 0.19$ pb predicted by MCFM. For the D0 measurements see~\cite{PRD-85-2012-112005}.
The ATLAS~\cite{arXiv:1208.1390} and CMS~\cite{CMS-PAS-EWK-11-010} measurements at 7 TeV
also agree with each other and with the SM predictions.
Study of WWZ ATGCs in 
$WZ \rightarrow \ell\nu\ell\ell$ production at D0~\cite{arXiv:1208.5458}
used
the reconstructed transverse momentum of the two 
leptons ($p_T^{\ell\ell}$) originating from the Z boson.
The effect of ATGCs is to increase the production cross
section, especially at high boson transverse momentum,
relative to its SM prediction, as shown in Fig.~\ref{fig:WWandWZ} (right). 
Summary of the limits on ATGCs measured in the WZ production can be found 
in~\cite{arXiv:1208.1390}. 

\subsection{ZZ production}

\begin{figure}[htbp]
\begin{center}
\includegraphics[width=0.43\textwidth]{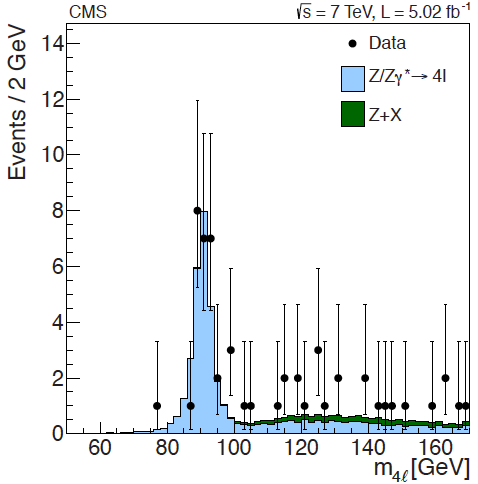}
\includegraphics[width=0.47\textwidth]{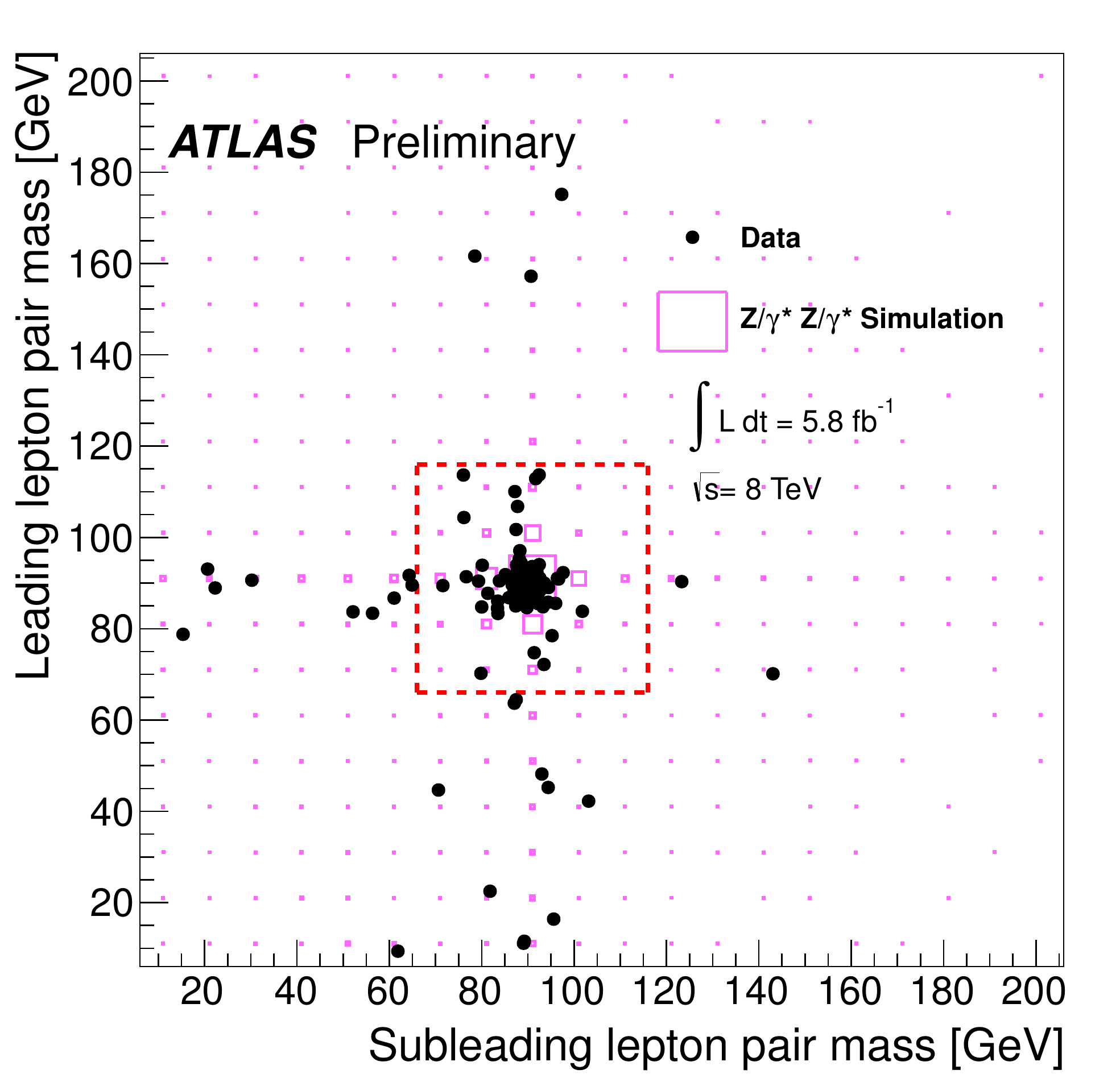}
\caption{ (left) 
Four-lepton mass distribution in $\rm{pp} \rightarrow \rm{Z} \rightarrow 4\ell$ decay channel. 
The data are shown by points. 
The filled histograms represent SM expectations for 
signal and
for reducible backgrounds, predicted using data. 
(right)
Invariant mass of the leading Z candidate versus the invariant mass of the subleading Z candidate in $\rm{pp} \rightarrow \rm{ZZ} \rightarrow 4\ell$ production. 
The events observed in 
the data are shown as solid circles and the signal prediction from simulation as pink boxes. The large dashed 
red box indicates the signal region defined by the ZZ fiducial cuts on the Z candidate masses. Contributions 
from events with one or both Z bosons off-shell are also seen.
\label{fig:ZZ1}}
\end{center}
\end{figure}

Production of the ZZ final states, which was always statistically limited 
at the Tevatron, profits at the LHC from both, increase in the luminosity and
in the center-of-mass energy.
Even if the main ZZ channel at the LHC remains the $ZZ \rightarrow 4\ell$,
ATLAS reported the cross section measurement using also $\rm{ZZ} \rightarrow \ell\ell\nu\nu$
decay channel~\cite{arXiv:1211.6096}. 
It is also important to mention, that working on the ZZ, CMS reported
a first observation of the Z boson decaying to four leptons in 
proton-proton collisions~\cite{arXiv:1210.3844}.
A pronounced resonance peak of 26 events was observed in the mass window 80-100 GeV
in the invariant mass distribution of four leptons with its mean and
width consistent with the Z boson, as shown in Fig.~\ref{fig:ZZ1}(left).
The product of the measured cross section and branching fraction, 
$\sigma \times BR(Z \rightarrow 4\ell) = 125 \pm 26(stat.) \pm 9(syst.) \pm 7(lumi.)$ fb, is consistent 
with the SM prediction of 126 fb.
Figure~\ref{fig:ZZ1}(right) shows
relationship between reconstructed dilepton masses of the highest-$p_T$ Z candidates 
and the trailing one in ${\rm ZZ} \rightarrow 4\ell$ decay channel with $\ell = e,\, \mu$~\cite{ATLAS-CONF-2012-090}. 
For the cross section measurements only events are used with 
$60(66) < m_{\rm{Z1}}, m_{\rm{Z2}} < 120(116)$ GeV for CMS(ATLAS).
\begin{figure}[htbp]
\begin{center}
\includegraphics[width=0.45\textwidth]{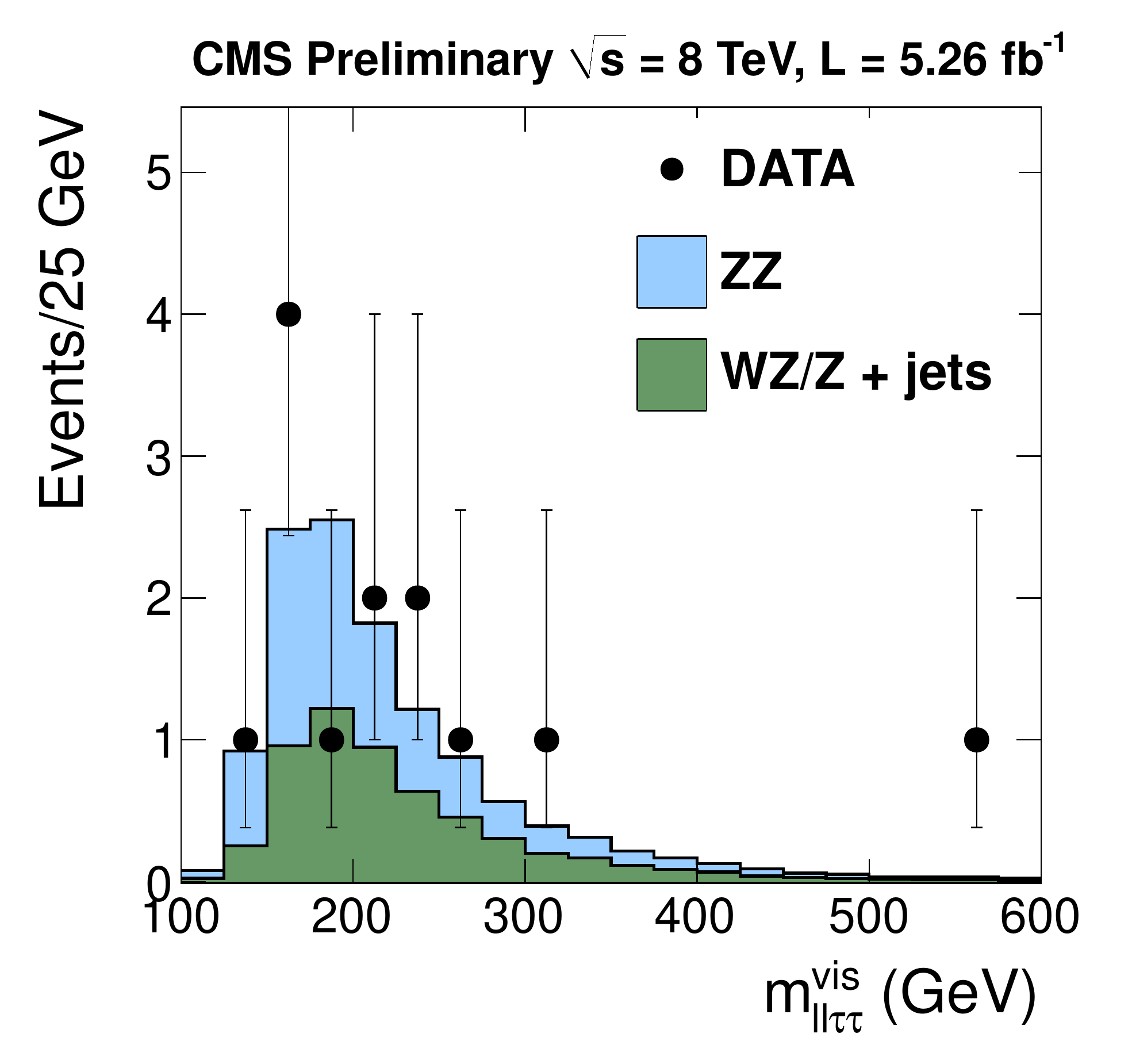}
\includegraphics[width=0.45\textwidth]{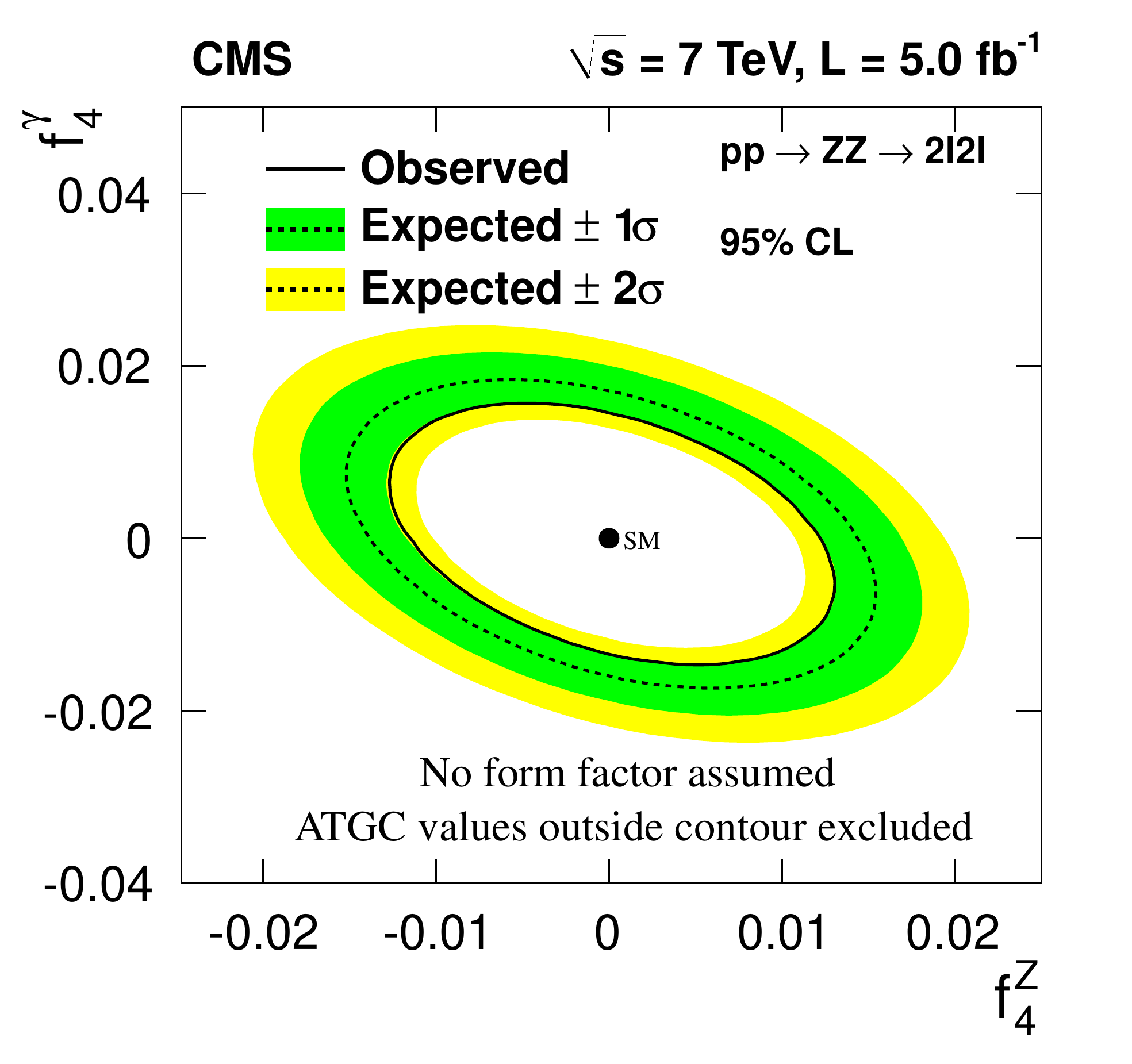}
\caption{ (left) 
Distributions of the four-lepton reconstructed mass for 
the sum of the $2\ell2\tau$ channels~\protect\cite{CMS-PAS-SMP-12-014}. 
Points represent the data, and the shaded
histograms represent the expected ZZ signal and the reducible background. The shapes of the
signal and background are taken from the MC simulation, with each component normalized to
the corresponding estimated value from data.
(right)
Expected and observed two-dimensional exclusion limits at 95\% CL on the anomalous
neutral trilinear ZZZ($f^{\rm{Z}}_{4}$) and ZZ$\gamma$ ($f^{\gamma}_4$) 
couplings~\protect\cite{arXiv:1211.4890}. The green and yellow bands represent the
one and two standard-deviation variations from the expected limit. In calculating the limits,
the anomalous couplings that are not shown in the figure are set to zero
\label{fig:ZZ2}}
\end{center}
\end{figure}
In the CMS paper the ZZ cross section was measured aslo using $\rm{ZZ} \rightarrow 2\ell2\tau$  decay mode.
The invariant mass of the $2\ell2\tau$ system is presented in Fig.~\ref{fig:ZZ2}(left). Finally
limits on ZZZ and ZZ$\gamma$  ATGCs are set as presented in Fig.~\ref{fig:ZZ2}(right).
All the recent CMS~\cite{arXiv:1211.4890,CMS-PAS-SMP-12-014} 
and ATLAS~\cite{arXiv:1211.6096,ATLAS-CONF-2012-090} measurements agree well with the SM predictions.


\section{Summary}

In this overview we summarized some of the recent measurements related to
the production of the W and Z bosons,
more information can be found in the references.
All experiments, both at the Tevatron and the LHC, 
performed cross section measurements of different channels, 
charged and forward-backward asymmetry measurements, set limits on ATGCs.
Since the cross sections are increasing with the center-of-mass energy, the LHC
has an opportunity to improve all the measurements, which are statistically
limited at the Tevatron. Especially in the diboson production, the LHC  
profits from both the increased production cross section and the integrated luminosity.
In the 2010-2012 LHC data taking periods 
the total integrated luminosity of proton-proton collisions is expected to reach
25--30 fb$^{-1}$.

The results of the measurements are generally well described by the SM
predictions. The
theoretical systematics play a major role in some of the measurements already.

\end{document}